# False Starts in History of Searches for 2β Decay, or Discoverless Double Beta Decay[1]

V.I. Tretyak

*Institute for Nuclear Research, MSP 03680, Kyiv, Ukraine*

**Abstract.** A collection of stories is presented on discoveries of 2β decay (including neutrinoless one) which were refuted in the subsequent investigations.

## INTRODUCTION

Analyzing the data of the Heidelberg-Moscow experiment, in 2001 H.V. Klapdor-Kleingrothaus with colleagues [1] concluded that "First evidence for neutrinoless double beta decay is observed … The evidence for this decay mode is 97% (2.2σ) …". However, in fact, it was not the first time in history of 2β searches when observation of the effect was declared. Already in 1966 V.R. Lazarenko wrote in his review: "Double beta decay was "discovered" many times, but all these discoveries were either refuted by further experiments, or were subjected to serious doubts for various reasons" [2]. In the following, I will list, in chronological order, all the previous declarations for observation of 2β decay (including neutrinoless!). In some cases authors were strongly sure that they really see the effect, in some other cases conclusions were not strong. It is worthwhile to remember these stories, and especially useful they could be for younger 2β physicists which are mostly unaware of them.

## WRONG 2β DECAY DISCOVERIES

<u>1949, $^{124}$Sn, 2β0ν evidence.</u> First experimental article on 2β decay appeared in 1948 [3]. It was a short note (a quarter of page) by E. Fireman who reported only the limit $T_{1/2}^{0\nu} > 3 \times 10^{15}$ yr for 2β0ν decay of $^{124}$Sn. But already in the second article [4] the 2β0ν effect was observed! Two samples of Sn, 25 g each, were used: sample A enriched in $^{124}$Sn to 54% (natural abundance is δ=5.8%) and sample B depleted in $^{124}$Sn to 0.4%. Two proportional counters were used for each Sn sample (working with and without coincidence); the samples were periodically rotated relatively to counters to exclude systematic effects. Difference in rates A−B was consistent with 0 without

---

coincidence (0.0±0.4 count/min) but in coincidence it was equal 2.0±0.2 count/min. Also, absorption curve with Al absorber was "similar to that of electrons from a spectrum with an energy end point between 1.0 Mev and 1.5 Mev" (in accordance with $^{124}$Sn $Q_{2\beta}$=2288 keV). Author concluded that "If one interprets this effect as double beta-decay from Sn$^{124}$, one obtains a half-life between $0.4\cdot10^{16}$ yr. and $0.9\cdot10^{16}$ yr. Other alternative explanations for these observations have been considered but none have been found to be plausible. This result would indicate that double beta-decay is unaccompanied by neutrinos". So, 2β0ν decay at the first time was "observed" in 1949 with 2.6σ significance (even better than 2.2σ observation in 2001 [1]). Currently known limit on $^{124}$Sn 2β0ν decay is $T_{1/2}^{0\nu}>2.0\times10^{19}$ yr [5], so what E. Fireman observed in 1949 was contamination of Sn sample A by some radioactivity.

1952, $^{100}$Mo, possible 2β effect. In photoemulsion experiment of J.H. Fremlin and M.C. Walters [6] which was, to our knowledge, the first underground (567 m deep) measurements in history of 2β searches, sixteen different samples were investigated. The photoemulsuion was sensitive to β and α particles. Excessive (on the level of 6–12σ) β activity was observed for Sn, $^{124}$Sn, Ba, W, Os samples – however, for all of them also α activity was detected. Thus, both of them were probably related with U/Th contanination. But for Mo excessive β activity (13σ) was not accompanied by α's, and this could be considered as indication on 2β decay of $^{100}$Mo (δ=9.8%) with $T_{1/2}=1.5\times10^{16}$ yr. No strong conclusion was made by the authors who noted that work on other samples is desirable. Today $T_{1/2}$ values for $^{100}$Mo 2β decay are: $T_{1/2}^{2\nu}=7.1\times10^{18}$ yr, $T_{1/2}^{0\nu}>4.6\times10^{23}$ yr [7], and most probably the effect was caused by some other (not U/Th) contamination of the Mo sample (f.e. by $^{90}$Sr/$^{90}$Y).

1953, $^{96}$Zr, possible 2β0ν effect. Two trans-stilbene scintillators (working independently or in coincidence) were applied by J.A. McCarthy to measure two samples of ZrO$_2$ (52 mg each): enriched in $^{96}$Zr to 89.5% (nat. δ=2.8%) and enriched in $^{94}$Zr to 97.9% (nat. δ=17.4%) [8]. Peak at ~3.8 MeV was observed (today value of $^{96}$Zr $Q_{2\beta}$ is 3348 keV), and difference in activity $^{96}$Zr–$^{94}$Zr was significant at ~3σ both in the total and coincidence spectra giving $T_{1/2}^{0\nu}=(6\pm2)\times10^{16}$ yr. The author wrote: "In view of the coincidence data … the author considers … double beta-decay … the most likely one". However, not a strong conclusion was given: "… The results presented here indicate, but do not prove, that double beta-decay may occur in Zr$^{96}$ without the emission of neutrinos. Further experimentation is clearly necessary". Today $T_{1/2}$ values for $^{96}$Zr are: $T_{1/2}^{2\nu}=2.4\times10^{19}$ yr [9], $T_{1/2}^{0\nu}>9.2\times10^{21}$ yr [10].

1955, $^{48}$Ca, 2β0ν evidence. Similar method J.A. McCarthy used also to measure two samples of CaCO$_3$: enriched in $^{48}$Ca to 84.3% (nat. δ=0.187%) and enriched in $^{44}$Ca to 97.9% (nat. δ=2.1%) [11]. Good peak was observed after 755 h in difference $^{48}$Ca–$^{44}$Ca coincidence spectrum (0.19±0.06 count/h) in the energy window 3.75–4.50 MeV ($^{48}$Ca $Q_{2\beta}$=4274 keV). Half-life was calculated as $T_{1/2}^{0\nu}=(1.6\pm0.7)\times10^{17}$ yr. This time the author gave strong conclusion: "The author believes this to be evidence for double beta decay without neutrino emission unless the observed counts are due to an unusual phenomenon of unknown origin". The today values are: $T_{1/2}^{2\nu}=4.4\times10^{19}$ yr [9], $T_{1/2}^{0\nu}>5.8\times10^{22}$ yr [12], so once more it was some radioactive contamination.

1980, $^{82}$Se, 2β2ν evidence. This is probably the most impressive example among all incorrect observations of 2β decay. M.K. Moe and D.D. Lowenthal investigated 13.75 g of Se (enriched in $^{82}$Se to 97%) in form of thin foils (5.6 mg/cm$^2$) using cloud chamber (with magnetic field 1 kG) and multiwire proportional counter [13]. It was possible to see tracks of two e$^-$ emitted simultaneously from one point (and distinguish them from e$^+$) and to measure energy of each e$^-$ and opening angle. The measurements were performed at the Earth surface under shielding of iron (>38 cm) and lead (15 cm). All materials were preliminary selected with NaI detector. In very detailed article (18 pages), all possible sources which could mimic 2β events were thoroughly analyzed. Twenty clean 2e$^-$ events were seen (5 were ascribed to $^{214}$Bi), and measured spectra for energies of single electrons, sum of their energies and opening angle were in good agreement with the expected ones for 2β2ν decay of $^{82}$Se. The obtained half life $T_{1/2}^{2\nu}=(1.0\pm0.4)\times10^{19}$ yr also was in agreement with some theoretical predictions. Clean observation of the effect (with tracks of particles), agreement of all spectra and $T_{1/2}$ with expectations – this was accomplished dream of experimentalist. Only one drawback existed: the obtained $T_{1/2}$ was inconsistent with the value known from geochemical experiments ~$10^{20}$ yr. Few years later, next measurements of S.R. Elliott, A.A. Hahn and M.K. Moe with new apparatus – TPC with magnetic field – gave $T_{1/2}^{2\nu}=(1.1^{+0.8}_{-0.3})\times10^{20}$ yr [14]. This work is considered as the first direct laboratory observation of 2β2ν decay; 35 events were detected during 7960 h. The today value (NEMO-3, ~1 kg of $^{82}$Se, 2750 events during 389 d) is $9.6\times10^{19}$ yr [7].

1987, $^{76}$Ge, 2β0ν decay with emission of Majoron. In measurements with HP Ge detector 135 cm$^3$ during 402 d in the Homestake gold mine, F.T. Avignone III with colleagues observed a bump resembling 2β0νM decay of $^{76}$Ge with $T_{1/2}^{0\nu M}=(6\pm1)\times10^{20}$ yr. This result was not published in a refereed journal but was reported at several conferences (f.e. [15]); conclusion was not strong: the authors wrote that they cannot claim at this point to have observed 2β0ν decay with emission of Majoron. However, it is interesting to note reaction of the press. The New York Times wrote: "Experiments conducted deep in a South Dakota gold mine have reportedly produced evidence for an extremely rare form of radioactive decay whose existence, if proved, would overthrow one of the basic laws of physics" [16]. It was noted in Nature: "As the existence of the majoron would represent a new particle which could not be accommodated in the standard model, these reports attracted great attention" [17]. Article in Science was titled: "Possible First Hints of Double Beta Decay" [18].

1988, $^{76}$Ge, 2β0ν decay to excited $2^+$ level of $^{76}$Se. In experiment, performed by the French (P. Mennrath et al.) and Spanish (A. Morales et al.) groups, 4 HP Ge detectors (~100 cm$^3$ each) worked in 4π active shielding of 19 NaI. Daughter $^{76}$Se nucleus has the first excited level at $E_{exc}(2_1^+)=559.1$ keV, and coincidence signal (559 keV in NaI and rest 1482 keV in Ge) was seen. After 6207 h measurements in the Modane Underground Laboratory, area of the coincident peak was 13±5 counts (2.6σ significance), the events were randomly distributed in time, and all Ge and NaI detectors were involved. The authors wrote: "All these features suggest the possibility of neutrinoless double beta decay of $^{76}$Ge to the first excited state of $^{76}$Se with a half-life $T_{1/2}=10^{22}$ y" [19]. The today limit, nevertheless, is 2 orders of magnitude higher: $>8.2\times10^{23}$ yr [20]. However, it was not the end of the story.

In paper [21], signed only by the Spanish group, the above results were presented with no strong conclusion ("... rather questionable to positively conclude that we are seeing a double beta effect"), and in the abstract only the limit >6×10$^{22}$ yr was given. Comments to this paper from the French group [22] started with words: "It cames as a real surprise to read in one of the latest issues of this journal" article [21]. J. Busto et al. wrote that, while both groups agree on existence of the small peak, there existed some differences of opinion on presentation of the results which had to be solved before common publication. "Apparently the group of Prof. A. Morales has preferred to abandon any further scientific discussion and to publish, as their own, the main results of a joint experiment. This group did not even send us a preprint of the article ... This action constitutes a serious breach of scientific ethics ..." [22]. In their reply [23], A. Morales et al. said that in fact it was the French colleagues who made the final decision without continuing further scientific discussion and insisting on publication of the results [19] as an indication of 2β decay: "... The title and the content of the transparencies presented by Prof. Mennrath at Heidelberg and previously agreed, were modified in the published version by our Bordeaux colleagues, without our knowledge nor our consent to sign the paper ...". Further details can be found in [19,21-23].

<u>1992-1994, 2β0ν decay of $^{130}$Te and background peak at 2527 keV.</u> In the data collected with 8 HP Ge detectors at the St. Gotthard Underground Laboratory during 15058 h [24], a small peak at 2527 keV was observed. This has extraordinary importance for calorimetric searches for 2β0ν decay of $^{130}$Te because its $Q_{2\beta}$ is also 2527 keV. Italian physicists, involved in $^{130}$Te experiment, wrote in [25]: "Due to relevance that this peak would have in the present and future experiment on $^{130}$Te, we have re-analized this spectrum in collaboration with the Neuchatel group ... We found unambiguously that a peak at 2527 keV was generated ... by the early saturation of amplifier number six, which was defective in the last part of the experiment". M. Moe and P. Vogel in their review [26] noted: "The existence of a background γ ray so close to the ... 0ν in $^{130}$Te would be a disaster for the tellurium bolometric experiments. After three months of intense effort to identify the peak, the Milan and St. Gotthard groups working together determined conclusively that the Gotthard feature was an artifact produced by a malfunctioning amplifier. With the nightmare of a 2527-keV γ ray behind them, the Milan group is proceeding with the tellurium experiments".

<u>1993, $^{82}$Se, $^{100}$Mo, $^{150}$Nd, indication of 2β0νM decays.</u> Thin foils of enriched 2β isotopes $^{82}$Se, $^{100}$Mo and $^{150}$Nd were measured with TPC in magnetic field during 20244, 6327 and 2534 h, respectively [27]. Two electron events at energies higher than the corresponding 2β2ν spectra were observed for all 3 isotopes. Rate of these events was higher for isotopes with lower $T_{1/2}^{2\nu}$; also their energies were higher for isotopes with higher $Q_{2\beta}$. However, the effect disappeared after increase of the strength of the magnetic field.

<u>1994, $^{76}$Ge, earlier indication of 2β0ν decay in the H-M experiment.</u> Dependence of the deduced $T_{1/2}^{0\nu}$ limit for $^{76}$Ge on time of measurements t in the Heidelberg-Moscow experiment was reported in [20]. It did not increase as $t^{1/2}$, as expected, but saturated at the value of 1.6×10$^{24}$ yr. However, no strong conclusion was given in [20].

<u>1995, $^{64}$Zn, indication of εβ$^+$ decay.</u> Zinc sample ⌀7×2.5 cm ($^{64}$Zn nat. δ=48.3%) was measured with HP Ge and NaI(Tl) detectors in coincidence [28]. Pure Fe and Cu cylinders served as blank samples. The $\varepsilon\beta^+$ should give two 511 keV $\gamma$'s emitted after annihilation of positron. They were detected with the Zn sample; corresponding half life is $T_{1/2}^{0\nu+2\nu}=(1.1\pm0.9)\times10^{19}$ yr. The conclusion was not strong: "If not caused by some unidentified and highly unlikely contamination of our Zn sample … this result will be the first experimental evidence for the electron positron conversion process". However, the today limits are: $T_{1/2}^{2\nu}>7.0\times10^{20}$ yr, $T_{1/2}^{0\nu}>4.3\times10^{20}$ yr [29].

<u>2008, $^{112}$Sn, possible indication of $\varepsilon\beta^+$ decay.</u> Excess of events in the 511 keV peak was also observed in measurements of 1.24 kg of $^{nat}$Sn with 72 cm$^3$ Ge detector [30]. The difference between number of events with and without Sn sample during 831 h was 270±50 (5.4$\sigma$ significance), and it was not related with U/Th contamination because of no excess for other U/Th peaks. The authors wrote: "Assuming the observed events are all due to the double beta decay of $^{112}$Sn, a half-life … can be derived" as $T_{1/2}^{0\nu+2\nu}=(5.1^{+1.2}_{-0.8})\times10^{17}$ yr. However, because an additional positron source in the tin cannot be ruled out, only limit $>4.1\times10^{17}$ yr was given finally. The best limit known today is $>9.7\times10^{19}$ yr [31].

History of searches for 2$\beta$ decay is a history of studies of backgrounds and fightings with them, and stories collected here once more demonstrate this. The claim [1] will be checked in forthcoming GERDA and Majorana experiments (see [32]).

I would like to thank the MEDEX'2011 organizing committee for support and warm atmosphere. I am thankful to S. Scopel for the second title of this paper.